\documentclass[fleqn,twoside]{article}
\usepackage[headings]{espcrc2}
\usepackage{epsf,epsfig}
\readRCS $Id: espcrc2.tex,v 1.2 2004/02/24 11:22:11 spepping Exp $
\usepackage{graphicx}
\usepackage[figuresright]{rotating}

\def\beq{\begin{eqnarray}}
\def\eeq{\end{eqnarray}}
\def\bea{\begin{eqnarray}}
\def\eea{\end{eqnarray}}\def\be{\begin{equation}}
\def\ee{\end{equation}}\def\nn{\nonumber}
\def\slash#1{#1 \hskip-0.50em /}

\newcommand{\AmS}{{\protect\the\textfont2
  A\kern-.1667em\lower.5ex\hbox{M}\kern-.125emS}}

\title{Light-cone sum rules: A SCET-based formulation}

\author{F. De Fazio\address{Istituto Nazionale di Fisica
Nucleare, Sezione di Bari, Italy}\thanks{Speaker},
        Th. Feldmann\address{Fachbereich Physik,
  Universit\"at Siegen, D-57068 Siegen, Germany}
        and
        T. Hurth\address{CERN, Dept.\ of Physics, Theory Division, CH-1211 Geneva 23, Switzerland
\\ and SLAC, Stanford University, Stanford, CA 94309, USA.}\thanks{Heisenberg Fellow}}

\begin{document}

\begin{abstract}
We describe the construction of light-cone sum rules (LCSRs) for
exclusive $B$\/-meson decays into light energetic hadrons from
correlation functions within soft-collinear effective theory
(SCET).  As an example, we consider the SCET sum rule for the $B
\to \pi$ transition form factor at large recoil, including
radiative corrections from hard-collinear loop diagrams at first
order in the strong coupling constant.  \vspace{1pc}
\end{abstract}

\maketitle

\section{INTRODUCTION}

Form factors parameterizing  hadronic matrix elements  defining
$B$ decays to a light pseudoscalar ($P$) or vector meson ($V$)
play an important role in several respects. For example, they
enter in the determination of $|V_{ub}|$ from exclusive modes.
However, since they are non-perturbative objects, their
determination is a difficult task.

Let us consider for definiteness the decay  of a $B$ meson in its
rest frame into a highly energetic pion.  Several energy scales
are involved: i) $\Lambda = {\rm few} \times \Lambda_{\rm QCD}$,
the {\it soft} scale set by the typical energies and momenta of
the light degrees of freedom in the hadronic bound states; ii)
$m_b$, the {\it hard}\/ scale set by the $b$\/-quark mass; iii)
the hard-collinear scale $\mu_{\rm hc}=\sqrt{m_b \Lambda}$
appearing via interactions between soft and energetic modes. The
dynamics of hard and hard-collinear modes can be described
perturbatively in the heavy-quark limit. The separation of the two
perturbative scales from the non-perturbative hadron dynamics is
formalized within the framework of soft-collinear effective theory
(SCET) \cite{Bauer:2000yr,Beneke:2002ph}. The small expansion
parameter in SCET is given by  $\lambda = \sqrt{\Lambda/m_b}$,
such that $ \lambda^2 m_b \ll \mu_{\rm hc} \sim \lambda m_b \ll
   m_b $.

 SCET describes $B$ decays to light hadrons with energies much
larger than their masses, assuming that their constituents have
momenta almost collinear to the hadron momentum $p^\mu$.
Introducing two light-like vectors $n_+^\mu=(1,0,0,-1)$,
$n_-^\mu=(1,0,0,1)$ one can generically write:
$p^\mu=p_+^\mu+p_-^\mu+p_\perp^\mu$, with $p_+^\mu=(n_-
p)/2n_+^\mu$, $p_-^\mu=(n_+  p)/2n_-^\mu$; momenta are then
classified according to the scaling of their light-cone
coordinates $(p_+,p_-,p_\perp)$.

In order to see how SCET can be exploited in the case of
heavy-to-light $B$ decays, we have to recall some general features
of the form factors relevant to these kind of transitions.

In the large energy limit of the  final state, $B \to P,V$ form
factors obey spin symmetry relations \cite{Charles:1998dr}, broken
by hard gluon corrections to the weak vertex and hard spectator
interactions. In the heavy-quark limit one can  write
\cite{Beneke:2003pa} (see also \cite{Lange:2003pk,Bauer:2002aj}):
\bea \langle \pi| \bar \psi \, \Gamma_i \, b |B\rangle =  C_i(E,
\mu_I) \,
\xi_\pi(\mu_I,E) &+  & \label{factorization} \\
   T_i(E,u,\omega,\mu_{\rm II}) \otimes \phi_+^B(\omega,\mu_{\rm II}) \otimes
         \phi_\pi(u,\mu_{\rm II})&+&
\nn   \ldots,
 \eea where $\Gamma$ is a generic Dirac
structure and the dots stand for sub-leading terms in
$\Lambda/m_b$. The matrix elements in (\ref{factorization}) get
therefore two contributions. The first one contains the
short-distance functions
 $C_i$,  arising from integrating
out hard modes: $\mu_I<m_b$, and a  ``soft'' form factor $\xi_\pi$
which   does not depend on the Dirac structure of the decay
current. In this contribution, the hard-collinear interactions are
        {\em not}\/ factorizable,
        so that  the ``soft''
        form factor is in general a non-perturbative object
        of order $(\alpha_s)^0$.
The second term in (\ref{factorization}) factorizes into a
        hard-scattering kernel $T_i$ and the light-cone distribution amplitudes
        $\phi_B$ and $\phi_\pi$.
$T_i$  contains the effect of both hard and hard-collinear
dynamics: $\mu_{\rm II}<\mu_{\rm hc}$. Both $C_i$ and $T_i$ can be
computed as perturbative series in $\alpha_s$, and potentially
large logarithms $\ln m_b/\mu_{\rm I}$ and $\ln \mu_{\rm
hc}/\mu_{\rm II}$ can be resummed by renormalization-group
techniques (the effective theories for the two short-distance
regimes are known as SCET$_{\rm I}$ and SCET$_{\rm II}$,
respectively). A still controversial  question is to what extent
the first contribution is numerically suppressed by Sudakov
effects.

Let us consider eq. (\ref{factorization}) for
$\Gamma_i=\gamma_\mu$, i.e. for the QCD vector current. This can
be matched onto SCET$_{\rm I}$ currents as follows
 \cite{Bauer:2000yr}:
 \be
  \bar q \, \gamma_\mu \, b  \rightarrow
  \left( C_4 n_-^\mu +C_5 v^\mu \right) {\bar \xi}_{\rm hc} W_{\rm hc} \,  Y_s^\dagger  h_v
 + \dots \,\,
\label{matching0} \ee where dots represent subleading terms and
$C_4=1 + {\cal O}(\alpha_s)$, $C_5 = {\cal O}(\alpha_s)$. $v^\mu$
is the heavy-quark velocity  with  $n_\pm v = 1$. The direction of
the momentum of the (massless) pion is given by $p_\pi^\mu =
(n_+p_\pi) \, n_-^\mu/2$. Besides, $\xi_{\rm hc}(x) = \frac{\slash
n_- \slash n_+}{4} \, \psi_{\rm hc}(x)$ is a hard-collinear
light-quark field in SCET$_{\rm I}$ and $h_v$ is the usual heavy
quark field in HQET. The hard-collinear and soft Wilson lines
$W_{\rm hc}$ and $Y_s$ appear to render the definition
gauge-invariant.

The soft form factor in (\ref{factorization}) can be defined as
\cite{Beneke:2003pa} \beq
 && \langle \pi(p') |
  (\bar \xi_{\rm hc} W_{\rm hc})(0) \, (Y_s^\dagger h_v)(0) |B(m_B v)\rangle
  =\nn \\
&&  (n_+p') \, \xi_\pi(n_+p', \, \mu_{\rm I}) \,,
\label{eq:softdef} \eeq

Neglecting  ${\cal O}(\alpha_s)$ effects the approximate symmetry
relations mentioned above between the vector and tensor form
factors for $B \to \pi$ transitions
 read  \cite{Charles:1998dr,Beneke:2000wa}: \bea
  f_+(q^2) &\simeq& \frac{m_B}{n_+p_\pi} \, f_0(q^2)
  \simeq \frac{m_B}{m_B+m_\pi} \, f_T(q^2) \nn \\
  &\simeq& \xi_\pi(q^2) \, .
\eea

SCET thus provides a field-theoretical framework to achieve the
factorization of short- and long-distance physics, and to
calculate the former in renormalization-group-improved
perturbation theory. However,  non-perturbative quantities such as
the soft form factors remain undetermined without further
phenomenological or theoretical input. A  theoretical approach for
this purpose is represented by  QCD/light-cone sum rules (see for
instance \cite{Khodjamirian,Ball,Colangelo:2000dp}). In
\cite{DeFazio:2005dx} we have shown  that it is possible to
formulate light-cone sum rules {\it within} SCET, in a different
way with respect to the traditional method. We summarize below the
main features of this new formulation.

\section{SUM RULES IN SCET: THE CASE OF  $B \to \pi$ DECAY}

In contrast to the traditional approach where the $B$ meson is
represented by an interpolating current,  we treat it as an
external field and not as a propagating particle in the
correlation function (see also \cite{Khodjamirian:2005ea}).
Actually,
 the heavy quark is nearly on-shell in the end-point
region. In SCET$_{\rm I}$ this is reflected by the fact that hard
sub-processes (virtualities of order $m_b^2$) are already
integrated out and appear in coefficient functions multiplying
$J_0$. Instead, the short-distance (off-shell) modes in SCET$_{\rm
I}$ are the hard-collinear quark and gluon fields. Hence, our
starting point is the correlator \beq
  \Pi(p') = i \int d^4x \,
   e^{i p'{} x} \, \langle 0 | T[ J_\pi(x) J_0(0)] | B(p_B) \rangle\,,
\label{piscet} \eeq where $p_B^\mu = m_B v^\mu$, and \bea
 J_0(0) &=& \bar \xi_{\rm hc}(0) W_{\rm hc}(0) Y_s^\dagger(0)
 h_v(0)\,,\label{J0} \\
  J_\pi(x) &\equiv& - i \, \bar \psi(x) \, \slash n_+ \gamma_5 \, \psi(x)
\nonumber \\[0.3em]
  & = & - i \, \bar \xi_{\rm hc}(x)  \,
          \slash n_+ \gamma_5 \, \xi_{\rm hc}(x) \label{eq:Jpiform} \\&&- i \,
\left( \bar \xi_{\rm hc} W_{\rm hc}(x)
       \slash n_+ \gamma_5 Y_s^\dagger q_s(x) + h.c.\right) \,,
\nn \eea where $q_s$ is the soft quark field in SCET and
  $\langle 0| J_\pi | \pi(p')\rangle = (n_+p') \, f_\pi$.
In the following we will consider a reference frame where
$p'_\perp=v_\perp=0$ and $n_+v = n_-v = 1$. In this frame the two
independent kinematic variables are $
 (n_+ p'{}) \simeq 2 E_\pi = {\cal O}(m_b) $, $
 0 > (n_- p'{}) = {\cal O}(\Lambda)$, with $|n_-p'| \gg m_\pi^2/(n_+p')$. The dispersive analysis
will be performed with respect to $(n_-p')$ for fixed values of
$(n_+p')$.

As with all QCD sum rule calculations, the procedure consists in
writing the correlator (\ref{piscet}) in two different ways: we
will refer to them as the  {\it hadronic} side and the {\it SCET}
side. On the hadronic side, one can write:

\begin{equation}
\Pi^{\rm HAD}(n_-p')= \Pi(n_- p') \Big|_{\rm res.}+ \Pi(n_- p')
\Big|_{\rm cont.} \,;
\end{equation}
the first term represents the contribution of the pion, while the
second takes into account the role of higher states and continuum
above an effective threshold $\omega_s = {\cal
O}(\Lambda^2/n_+p')$.
One has \beq
  \Pi(n_- p') \Big|_{\rm res.} &= &
   \frac{\langle 0|J_\pi |\pi(p')\rangle \langle \pi(p')|J_0|B(p_B)\rangle}
    {m_\pi^2-p'{}^2} \nn \\
 & = &
 -{(n_+p') \xi_\pi(n_+ p')f_\pi \over n_- p'}\,,
\eeq obtained  in the chosen frame where $p'_\perp=0$ and
neglecting the pion mass.
\begin{figure}[htb]
\vspace{-0.6cm}\begin{center} \psfig{file=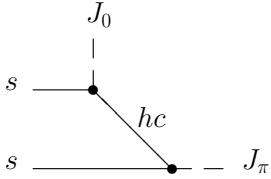,
width=0.22\textwidth} \caption{Leading contribution to the
correlation function for the
 SCET current $J_0$.} \end{center}\label{diag-a}
\end{figure}
At tree level, the SCET side stems from calculating the diagram in
Fig.~1, with the result: \beq
  \Pi(n_- p'{}) & =  &
  f_B m_B \int_0^\infty d\omega \,
   \frac{\phi_-^B(\omega)}{\omega - n_-p'{}-i\eta} \,,
\label{tmp} \eeq \noindent where $\omega=n_- \cdot k$, $k^\mu$
being the momentum of the soft light quark that ends up as
spectator in the $B$. In (\ref{tmp}) we  used the momentum-space
representation of LCDAs for $B$\/ mesons as in
\cite{Grozin:1996pq,Beneke:2000wa}, \bea
 && {\cal M}^B_{\beta\alpha} = - \frac{i f_B m_B}{4} \times\nn \\
 &&  \left[ \frac{1+\slash v}{2} \left\{ \phi_+^B(\omega) \slash n_+ +
   \phi_-^B(\omega) \slash n_- + \ldots \right\} \gamma_5
\right]_{\beta\alpha} \nn  . \eea

\begin{figure}[htb]
\vspace{-0.6cm} \psfig{file=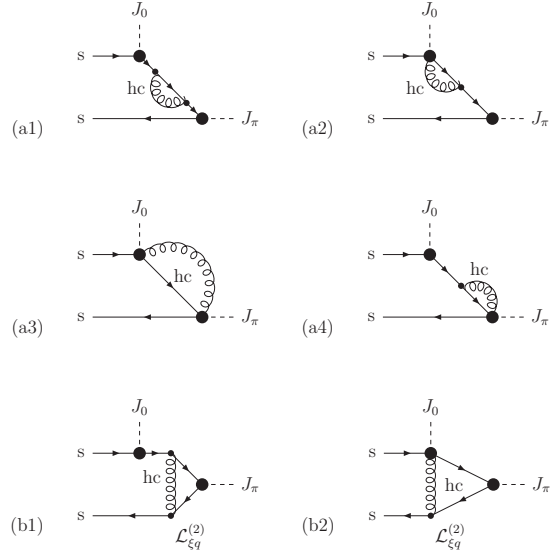, width=0.45\textwidth}
\caption{Diagrams contributing to the sum rule for $\xi_\pi$ to
order $\alpha_s$ with hard-collinear loops and no external soft
gluons.} \label{fig:hcloops}
\end{figure}

Notice that (\ref{tmp}) has already the form of a dispersion
relation in the variable $n_-  p^\prime$: \beq
  \Pi(n_- p')
  &=& \frac{1}{\pi} \int_0^\infty d\omega' \,
      \frac{{\rm Im}[\Pi(\omega')]}{\omega'-n_-p' - i\eta}\,,
\label{disp} \eeq with $\displaystyle{1 \over \pi}{\rm
Im}[\Pi(\omega^\prime)]=f_B m_B \phi_-^B(\omega^\prime)$. The
final sum rule  is obtained by writing also $\Pi(n_- p')
\Big|_{\rm cont.}$ according to a dispersion relation in which the
spectral function is identified  with the one computed in SCET.
Finally, a Borel transformation with parameter $\omega_M$ is
applied to both sides, giving the following sum rule at tree
level: \be
  \xi_\pi(n_+p') =  \frac{f_B m_B}{f_\pi (n_+p')} \,
   \int_0^{\omega_s} d\omega \, e^{-\omega/\omega_M} \, \phi_-^B(\omega)
 \,.
\label{sumexact} \ee The inclusion of radiative corrections to the
correlation function (\ref{piscet}) comes from  hard-collinear
loops, as shown in Fig.~\ref{fig:hcloops} for the leading order in
$\alpha_s$. The explicit calculation shows that the
scale-dependence of the correlation function cancels with that of
the $C_i(\mu)$ at the considered leading logarithmic order
(involving double logs).
As for the numerical analysis, we fix   the sum rule parameters
$\omega_S$ and $\omega_M$ from the sum rule for $f_\pi$, which
provides us with the {\it default values} $\omega_M \simeq
\omega_S \simeq 0.2 \,\,$ GeV. For $\phi_-^B(\omega)$, we use the
parametrization proposed in \cite{Grozin:1996pq}:
$\phi_-^B(\omega)= e^{-\omega/\omega_0}/\omega_0$ with $1/\omega_0
= \phi_-^B(0) = 2.15$~GeV$^{-1}$. Fixing one of the two parameters
to its default value and varying the other, we may investigate the
dependence on such quantities. It turns out that going from LO to
NLO such a dependence becomes moderate (see ref.
\cite{DeFazio:2005dx} for details). Taking into account the
various uncertainties, we obtain: \be \frac{C_i(\mu)}{C_i(m_b)}
\cdot  \xi_\pi(m_B,\mu) =
  0.27^{+0.09}_{-0.11}  \,,
\ee which compares well with other estimates for the $B \to \pi$
form factor in {\em full}\/ QCD.

Our approach can also be applied to calculate the factorizable
form factor contribution, which comes from spectator scattering
terms. This can be obtained starting from the correlator: \bea
  \Pi_1(p') = i \int d^4x \,
   e^{i p'{} x} \, \langle 0 | T[ J_\pi(x) J_1(0)] | B(p_B) \rangle \,,
\nn \eea where $
 J_1 =
 \bar \xi_{\rm hc} \,  g \, \slash A^\perp_{\rm hc} \,  h_v $ in the light-cone gauge.

The remarkable result of the SCET-sum-rule for the $B \to \pi$
form factor is that the ratio of factorizable and non-factorizable
contributions is independent of the $B$\/-meson wave function to
first approximation and amounts numerically to about $
   \approx 6 \% \,,$
 which is in line with the power counting used
in QCD factorization \cite{Beneke:1999br,Beneke:2000wa}, but
contradicts the assumptions of the  pQCD approach
\cite{Chen:2001pr} and the results of a recent study in
\cite{Bauer:2004tj}.

\section{CONCLUSIONS}

We have described the approach derived in \cite{DeFazio:2005dx}
consisting in  the derivation of  light-cone sum rules for
exclusive $B$\/-decay amplitudes at large recoil within
soft-collinear effective theory (SCET). This formalism defines a
consistent scheme to calculate both factorizable and
non-factorizable contributions to exclusive $B$ decays as a power
expansion in $\Lambda/m_b$. The non-perturbative information is
encoded in the light-cone wave functions of the $B$ meson, and in
the sum-rule parameters.

An explicit example is provided by the study of the factorizable
and non-factorizable contributions to the $B \to \pi$ form factor
at leading power in $\Lambda/m_b$.  The result for the central
value of the ``soft''/non-factorizable $B \to \pi$ form factor is
consistent with corresponding estimates  in full QCD. In
particular,  to first approximation, the {\em ratio}\/ of
factorizable and non-factorizable contributions is independent of
the $B$\/-meson wave function and small (formally of order
$\alpha_s$ at the hard-collinear scale, numerically of the order
of 5-10\%),  thus confirming the power-counting adopted in the
QCD-factorization approach.

The improvement of the SCET sum rule for the $B \to \pi$ form
factor and the extension to other  decays requires a better
understanding of both, the size and the renormalization-group
behaviour, of the light-cone wave functions for higher Fock states
in the $B$ meson. These issues are left for future investigations.

\end{document}